\documentclass{optica-article}

\journal{opticajournal} % for journals or Optica Open

\articletype{Research Article}

\usepackage{lineno}
\usepackage[normalem]{ulem}
\usepackage{multirow} % Required for \multirow command
\usepackage[table]{xcolor} % Required for table and color packages
\usepackage{comment}
\begin{document}

\title{Ultrabroadband tunable difference frequency generation in standardized thin-film lithium niobate platform}

\author{Yesim Koyaz,\authormark{1} Christian Lafforgue,\authormark{1} Homa Zarebidaki,\authormark{2} Olivia Hefti,\authormark{1,2} Davide Grassani,\authormark{2} Hamed Sattari,\authormark{2} and Camille-Sophie Brès\authormark{1,*}}

\address{\authormark{1}Ecole Polytechnique Fédérale de Lausanne, Photonic Systems Laboratory (PHOSL), STI-IEM, Station 11, CH-1015 Lausanne, Switzerland\\
\authormark{2}CSEM Neuchâtel, Rue Jaquet-Droz 1, 2002 Neuchâtel, Switzerland}

\email{\authormark{*}camille.bres@epfl.ch} %% email address is required; see note below about the corresponding author designation

% use {asbstract*} to suppress the copyright line. Copyright information will be added in production

\begin{abstract*} 
Thin-film lithium niobate (TFLN) on insulator is a promising platform for nonlinear photonic integrated circuits (PICs) due to its strong light confinement, high second-order nonlinearity, and flexible quasi-phase matching for three-wave mixing processes via periodic poling. Among the three-wave mixing processes of interest, difference frequency generation~(DFG) can produce long wave infrared~(IR) light from readily available near IR inputs. While broadband DFG is well studied for mid-IR frequencies, achieving broadband idler generation within the telecom window~(near C-band) and the short-wave infrared~(near 2 micron) is more challenging due to stringent dispersion profile requirements, especially when using standardized TFLN thicknesses. In this paper, we investigate various standard waveguide designs to pinpoint favorable conditions for broadband DFG operation covering several telecom bands.
Our simulations identify viable designs with a possible 3-dB conversion efficiency bandwidth (CE-BW) of 300~nm and our measurements show idler generation from 1418~nm to 1740~nm, limited by our available sources, experimentally confirming our design approach. Furthermore, temperature tuning allows further shift of the idler towards the mid-IR, up to 1819~nm. We also achieve a stretched wavelength range of idler generation by leveraging the longitudinal variation of the waveguide in addition to poling. Finally, our numerical simulations show the possibility of extending the CE-BW up to 780~nm while focusing on waveguide cross-sections that are available for fabrication within a foundry. Our work provides a methodology that bridges the deviations between fabricated and designed cross-sections, paving a way for standardized broadband DFG building blocks.
\end{abstract*}

%%%%%%%%%%%%%%%%%%%%%%%%%%  body  %%%%%%%%%%%%%%%%%%%%%%%%%%
\section{Introduction}
Many photonic integrated circuit~(PIC) platforms, such as silicon~(Si) or silicon nitride~(SiN), lack intrinsic second-order susceptibility~($\chi^{(2)}$), relying instead on third-order susceptibility~($\chi^{(3)}$) for nonlinear frequency conversion. However, $\chi^{(2)}$ enables nonlinear responses even at moderate input powers. Lithium niobate~(LN), which exhibits high $\chi^{(2)}$~($d_{33}\approx$~-25.2 pm/V where $\chi^{(2)}_{\textnormal{eee}}$~=~2$d_{33}$ \cite{Liu2005NonlinearDevices}) emerges as a promising PIC platform. Thin-film LN~(TFLN), available as a single crystal LN layer over the entire handling silicon-on-insulator wafer\cite{Nassau1966FerroelectricCrystals, Nassau1966FerroelectricEtching}, combined with advances in etching techniques\cite{Wang2017SecondWaveguides, Wu2018LongRoughness}, allows for low-loss waveguides with strong light confinement. Furthermore, LN's ferroelectric nature enables microscale crystal orientation control by applying high-voltage ms-pulses from periodic metal electrodes, whose periodicity,  precisely inscribed by lithography, becomes a design parameter\cite{Zhu2021IntegratedNiobate}. This technique, i.e. periodic poling, offers versatile means to achieve quasi phase-matching~(QPM) conditions making TFLN particularly favorable for nonlinear frequency generation.

This work focuses on difference-frequency generation~(DFG) using periodically-poled TFLN waveguides. DFG can carry out optical parametric oscillation~(OPO) and amplification~(OPA), facilitating optical processing and coherent amplification \cite{Cerullo2003UltrafastAmplifiers, Zhang2019OpticalAmplification}. Broadband idler generation, combined with advancements in extending the bandwidth of optical amplifiers (e.g., super-L, S, or O bands), offers a promising solution for achieving high-capacity data transmission and supporting advanced quantum modulation schemes. Moreover, DFG can support broadband generation of frequencies that are challenging to reach using conventional lasing materials. It was demonstrated for mid-IR frequencies covering idler wavelengths from 2.6~$\mu$m to 4~$\mu$m in TFLN on sapphire substrates at high temperatures~(>175°C) \cite{Mishra2022Ultra-broadbandNiobate} and for signal wavelengths around 2~$\mu$m (with a bandwidth $\sim$600~nm) \cite{Ledezma2022IntenseWaveguides}. Near the C-band transmission window (around 1550~nm), broadband idler generation has proved to be challenging due to more stringent restrictions in terms of achievable dispersion profiles. Nevertheless, DFG idler bandwidths up to 100~nm were demonstrated experimentally \cite{Niu2020OptimizingDomains, Sua2018Ultra-widebandWaveguide, Kishimoto2016HighlyWaveguide, Wei2024EfficientWaveguides}. Furthermore, spontaneous idler generation without requiring the signal input, i.e. spontaneous parametric down conversion~(SPDC), is also investigated for broadband response with potential applications in quantum signal processing \cite{Fang2024EfficientNanowaveguides, Javid2021UltrabroadbandChip, Xue2021UltrabrightChip}.

In this work, we design periodically poled waveguides for broadband DFG at telecom wavelengths, and specifically identify favourable cross-sections on a standardized TFLN platform, with a vision to incorporate nonlinear building blocks in the TFLN process design kit (PDK). We theoretically calculate the QPM DFG poling period, conversion efficiency (CE), and bandwidth. We experimentally validate one of our designs by measuring the broadband idler generation over a $\sim$300~nm wide wavelength range covering several telecom bands and demonstrate enhanced bandwidth towards mid-IR via temperature tuning. We also investigate the effect of longitudinal waveguide fluctuations by comparing the nonlinear response of two different waveguides and leverage this inhomogeneity for broadband operation.

\section{Dispersion Engineering and Design}
\label{sec2:disp}

Our analysis focuses on QPM DFG between fundamental TE modes of pump, signal and idler on a waveguide assuming perfect square poling. We perform numerical mode simulations~(in Ansys Lumerical) and calculate poling period~($\Lambda$), and CE at a selected poling period using Eq.~\ref{eq-poling-period} and Eq.~\ref{eq-ce} \cite{Suhara2003WaveguideDevices, Sutherland2003HandbookOptics}:

\begin{equation}
\label{eq-poling-period}
\frac{1} {\Lambda} = \frac{n_\textnormal{eff,s}}{\lambda_\textnormal{s}} + \frac{n_\textnormal{eff,i}}{\lambda_\textnormal{i}} - \frac{n_\textnormal{eff,p}}{\lambda_\textnormal{p}} 
\end{equation}

\begin{equation}
\label{eq-ce}
CE(z)= \frac{P_\textnormal{i}}{P_\textnormal{s} P_\textnormal{p}} = \frac{4}{\pi^2} \left|\gamma_\textnormal{DFG}\right|^2 z^2 \operatorname{sinc}\left(\frac{\Delta \beta}{2} z\right)^2
\end{equation}
where $n_\textnormal{eff, s/p/i}$, $\lambda_\textnormal{s/p/i}$ and $P_\textnormal{s/p/i}$ are effective index, wavelength and power of signal/pump/idler, respectively,  and $z$ is the periodically poled waveguide length. In the following analysis (throughout Sec.~\ref{sec2:disp}), we present the simulation results for $z$~=~5~mm.

The terms $\left|\gamma_\textnormal{DFG}\right|^2$ and $\Delta \beta$ correspond to nonlinear mode overlap~(Eq.~\ref{eq-nonlinearmodeoverlap}) and phase mismatch~(Eq.~\ref{eq-beta}) \cite{Suhara2003WaveguideDevices}:

\begin{equation}
\label{eq-nonlinearmodeoverlap}
\left|\gamma_\textnormal{DFG}\right|^2=\frac{\omega_\textnormal{i}^2\left(\chi^{(2)}\right)^2}{2 \varepsilon_o c^3 n_\textnormal{eff,i} n_\textnormal{eff,s} n_\textnormal{eff,p}} \frac{\left(\iint E_\textnormal{s}^*\textnormal{(x,y)} E_\textnormal{p}^*\textnormal{(x,y)} E_\textnormal{i}\textnormal{(x,y)dxdy}\right)^2}{
\displaystyle \prod_\textnormal{k=i,s,p} \iint\left|E_\textnormal{k}\textnormal{(x,y)}\right|^2 \textnormal{dxdy}} 
\end{equation}

\begin{equation}
\label{eq-beta}
\Delta \beta = \frac{n_\textnormal{eff,s}}{\lambda_\textnormal{s}} + \frac{n_\textnormal{eff,i}}{\lambda_\textnormal{i}} - \frac{n_\textnormal{eff,p}}{\lambda_\textnormal{p}} -\frac{2 \pi}{\Lambda}
\end{equation}
where $E_\textnormal{s/p/i}$ is the signal/pump/idler electric field.

\begin{figure}  [!htb]
\centering\includegraphics[width=13.5cm]{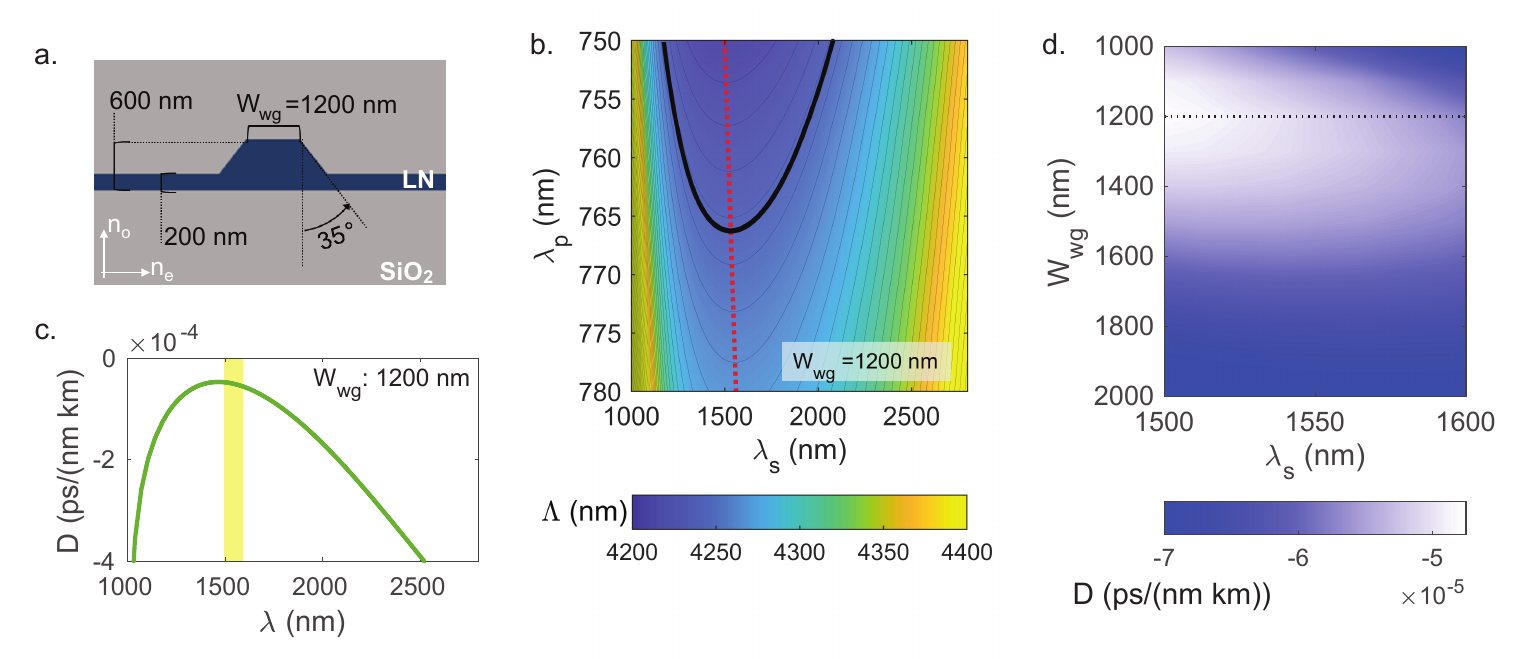}
\caption{ a. Schematic of investigated X-cut TFLN waveguide cross-section with SiO$_{2}$ top cladding. b. Poling period $\Lambda$ calculated for QPM DFG for the cross-section shown in (a) while sweeping pump~($\lambda_{p}$) and signal~($\lambda_{s}$). $\Lambda$~=~4241~nm is shown in black line and SH-line is shown in red line. c) Group velocity dispersion~(D) of the investigated cross-section as shown in (a). The region corresponding to $\lambda_{s}$ in degenerate DFG~($\lambda_{s}=\lambda_{i}$) for $\lambda_{p}$ ranging from 750~nm to 800~nm is highlighted in yellow.  d) D for varying waveguide width $W_{wg}$ and within the wavelength region of interest.}
\label{Fig1}
\end{figure}

We first engineer the DFG response via $\Lambda$~(Fig.~\ref{Fig1}b) calculated for different $\lambda_{s}$ and $\lambda_{p}$ with the waveguide cross-section given in Fig.~\ref{Fig1}a, compatible to the TFLN PIC foundry of CSEM\cite{Sattari2024StandardizedPlatformb} . For each $\Lambda$, two distant phase matched regions merge as $\lambda_{p}$ approaches the second-harmonic~(SH) wavelength~($\lambda_{p}\approx2\lambda_{s}$), which also corresponds to near-degenerate DFG~($\lambda_{s}\approx\lambda_{i}$). This wavelength relation is labelled in Fig.~\ref{Fig1}b as the red dashed line and referred as SH-line hereafter. We observe that $\Lambda$ remains relatively constant near the SH-line and as a result, this region becomes favorable for broadband operation. 

In this paper, we focus on optimizing the broadband response while sweeping $\lambda_{s}$ at a fixed $\lambda_p$. For the $\lambda_{s}$-sweep, conversion efficiency 3 dB bandwidth~(CE-BW) is conventionally associated with the group velocity mismatch (GVM) between signal and idler. Yet, near the SH-line, GVM approaches zero as $\lambda_{s}\approx\lambda_{i}$. As a result, higher order dispersion terms like the group velocity dispersion (D) starts to influence the bandwidth. For illustration, we plot in Fig.~\ref{Fig1}c, D for the cross-section given in Fig.~\ref{Fig1}a where the signal wavelength range of interest is highlighted.

\begin{figure}  [!htb]
\centering\includegraphics[width=13.5cm]{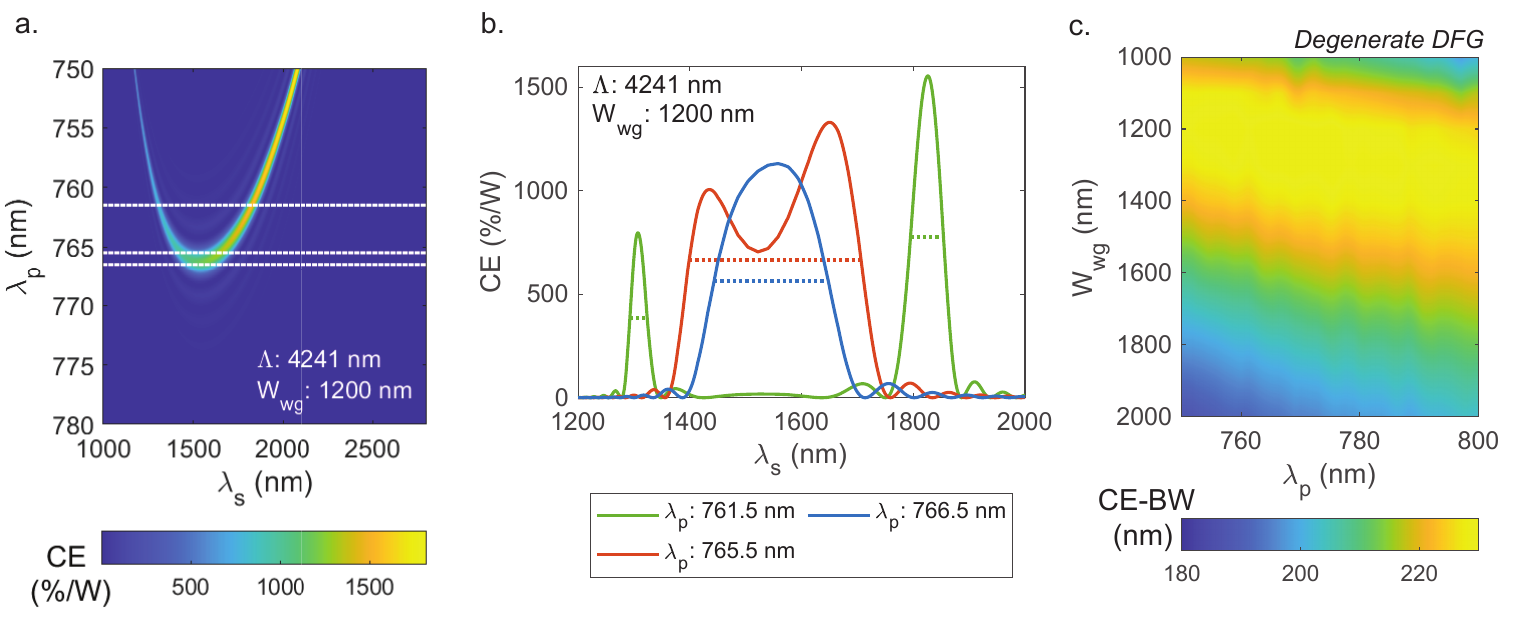}
\caption{a. Simulated QPM DFG conversion efficiency (CE) at $\Lambda$~=~4241~nm for the cross-section shown in Fig.~\ref{Fig1}a. b. CE plotted at the same $W_{wg}$ and $\Lambda$ for selected $\lambda_{p}$ namely 766.5~nm~($\lambda_{p}$ that satisfies degenerate DFG at this $\Lambda$), 765.5~nm~($\lambda_{p}$ that is optimized for broadband operation) and 761.5~nm~(a $\lambda_{p}$ that facilitates two separate phase-matched regions). Each case is labelled in white dashed lines in (a) and CE-BW of each peak is shown in dashed lines. c. CE-BW for different $\lambda_{p}$ and $W_{wg}$ where $\Lambda$ is selected to satisfy QPM for degenerate DFG on each data point.}
\label{Fig2}
\end{figure}

With the motivation of maximizing CE-BW, we extend the design space, considering varying waveguide widths~($W_{wg}$) while keeping the etch angle and the TFLN thickness constant, hence maintaining compatibility with the standardized platform. We numerically calculate D for each $W_{wg}$ ranging from 1000~nm to 2000~nm and for the same wavelength range highlighted in Fig.~\ref{Fig1}c.

Our simulations indicate that minimum magnitude of D~($|\textnormal{D}|$) occurs between $W_{wg}$~=~1100~nm and $W_{wg}$~=~1400~nm~(Fig.~\ref{Fig1}d). In literature, waveguides with near-zero D are associated with broad bandwidth and favored for different nonlinear processes like SPDC and four-wave mixing~(FWM)~\cite{Javid2021UltrabroadbandChip, Pu2018UltraEfficientProcessing}. The presented cross-section~(Fig.~\ref{Fig1}a) with $W_{wg}$~=~1200~nm lies in this favorable region. Therefore, we continue the analysis in the same cross-section by calculating CE~(Eq.~\ref{eq-ce}). The CE at a selected $\Lambda$ of 4241~nm, which is labelled in black on Fig.~\ref{Fig1}b and is phase-matched for degenerate DFG ($\lambda_{s}=\lambda_{i}$) at $\lambda_{p}\approx$~766.5~nm, is shown in Fig.~\ref{Fig2}a. The CE for a few selected pump wavelengths $\lambda_{p}$ is plotted in Fig.~\ref{Fig2}b. We see that for $\lambda_{p}$~=~766.5~nm (blue line) we obtain one broadband phase-matched peak. Then, as $\lambda_{p}$ is slightly detuned, the two phase-matched regions start separating. For $\lambda_{p}\approx$~765.5nm (orange line), we obtain the broadest CE-BW estimated at $\sim$307~nm and idler generation from $\sim$1357~nm to $\sim$1758~nm. We also observe the complete separation of the phase matched regions for shorter $\lambda_{p}$. For example, CE(@$\lambda_{p}$~=~761.5~nm) has two different phase-matched regions present at $\lambda_{s}$ around 1307~nm and 1827~nm with CE-BW of 29~nm and 59~nm, respectively (green line). Overall, we show tunable CE-BW from 29~nm to 307~nm via adjusting $\lambda_{p}$.

\begin{figure} [!htb]
\centering\includegraphics[width=13.5cm]{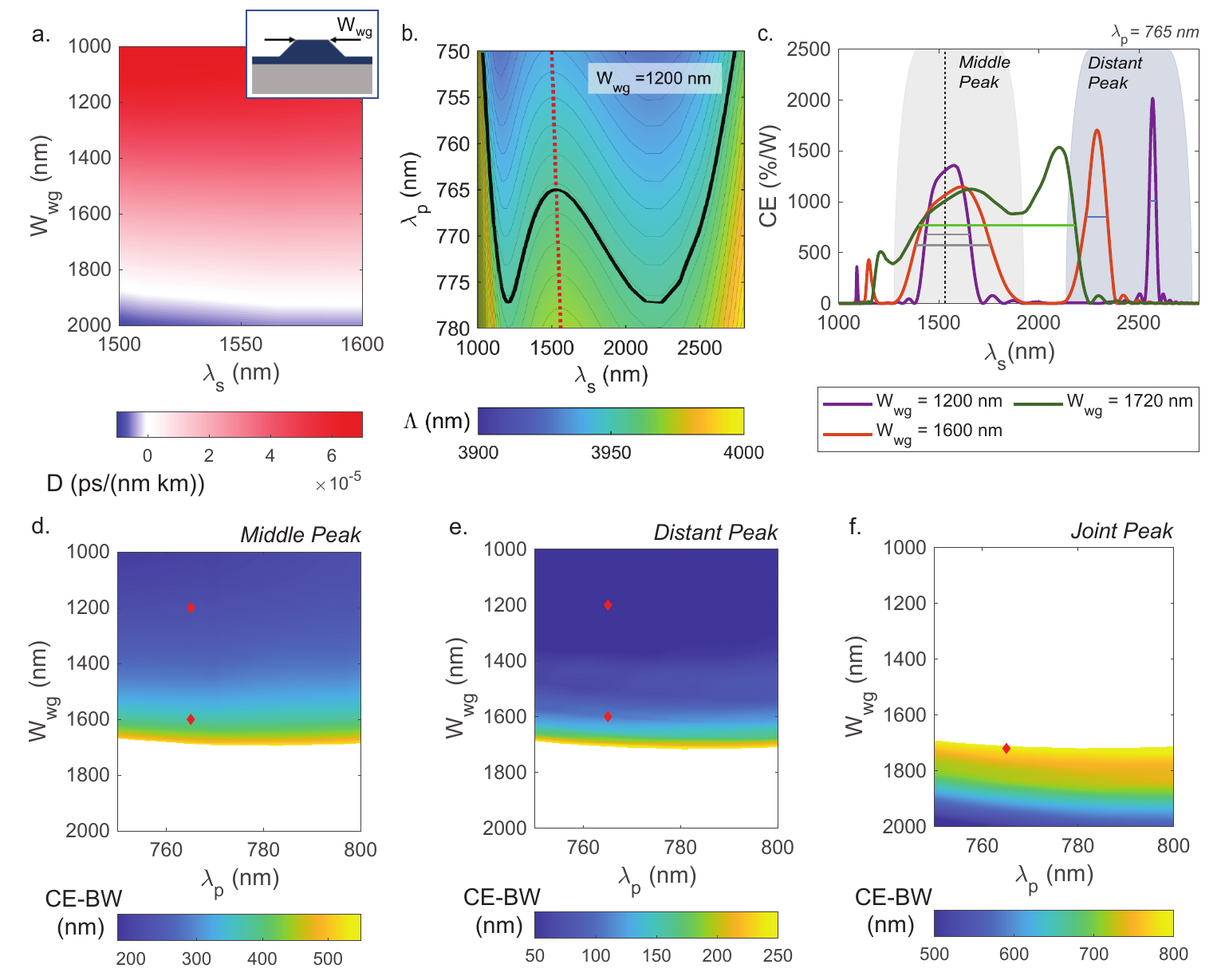}
\caption{DFG simulations for the uncladded waveguide cross-sections.  a. D for different $\lambda_{s}$ and $W_{wg}$ on the region that corresponds to degenerate DFG for $\lambda_{p}$ from 750~nm to 800~nm. b. $\Lambda$ calculated for $W_{wg}$~=~1200~nm while sweeping $\lambda_{p}$ and $\lambda_{s}$. SH-line is shown in red dashed line. $\Lambda$ optimized for degenerate DFG at $\lambda_p$~=~765~nm is shown in black line. c-f. Analysis for different $W_{wg}$ when $\Lambda$ is optimized to satisfy QPM of degenerate DFG. c. CE at $\lambda_{p}$~=~765~nm. $\lambda_{s}$ satisfying degenerate DFG is shown in black dashed line. CE-BW calculated for middle/distant/joint peaks are shown in grey/blue/green lines respectively and labelled in red markers in (d-f). d-f. CE-BW calculated for d.~middle~/~e.~distant~/~f.~joint peak at different $\lambda_{p}$ and $W_{wg}$.}
\label{Fig3}
\end{figure}

We also extend the CE calculations to different cladded cross-sections at the optimized $\Lambda$ to operate at degenerate DFG regime~($\lambda_{s}=\lambda_{i}$). In Fig.~\ref{Fig2}c, we plot CE-BW for different $W_{wg}$ ranging from 1000~nm to 2000~nm for $\lambda_{p}$ ranging from 750~nm to 800~nm. D for all $\lambda_{s}$ corresponding to the described operating point is already presented in Fig.~\ref{Fig1}c. In this design space, we indeed see that the maximum CE-BW occurs on minimum $|\textnormal{D}|$, verifying the proposed link between CE-BW and D in near degenerate DFG regime.

Next, we repeat the analysis for an uncladded configuration, which gives an extra degree of freedom, in addition to the waveguide dimensions, in terms of dispersion engineering. In Fig.~\ref{Fig3}a, we plot D at different $W_{wg}$ and $\lambda_s$. We observe that most of the investigated waveguides are in the anomalous dispersion regime in contrast to previous configuration. Due to this difference, we start from one $W_{wg}$~(namely 1200~nm which has strong anomalous dispersion) and calculate $\Lambda$ at different $\lambda_s$ and $\lambda_p$~(Fig.~\ref{Fig3}b). We observe that, when the $\Lambda$ is phase-matched to satisfy degenerate DFG~($\lambda_{s}$~=~$\lambda_{i}$), it is also phase-matched for a distant signal/idler pair at the same $\lambda_{p}$. As an illustration, we carry the following analysis for $\lambda_p$~=~765~nm and with the corresponding $\Lambda$ highlighted in black in Fig.~\ref{Fig3}b. The calculated CE is shown in purple in Fig.~\ref{Fig3}c. It is characterized by three phase-matched regions located at $\lambda_{s}$~=~$\sim$1089~nm, $\lambda_{s}$~=~$\sim$1530~nm and $\lambda_{s}$~=~$\sim$2568~nm) where the middle one fulfills the degenerate DFG condition. The peaks of interest referred as "middle peak" and "distant peak" are labelled in Fig.~\ref{Fig3}c. 

The influence of dispersion on this three phase matched regions is carried by extending the CE calculations for different $W_{wg}$ and is also illustrated in Fig.~\ref{Fig3}c. As $W_{wg}$ increases, D can remain anomalous but becomes smaller (see Fig.~\ref{Fig3}a). For example, for $W_{wg}$~=~1600~nm~(orange line), we observe that the CE-BW of the "middle peak" increases while the "distant peak" shifts towards the "middle peak". When $W_{wg}$ is further increased, D approaches to near-zero dispersion and the phase-matched bands starts to overlap. Particularly, at $W_{wg}$~=~1720~nm~(green line), we obtain a continuous range of idler generation from $\sim$1160~nm to $\sim$2253~nm and a CE-BW of $\sim$782~nm. 

We further extend the design space and calculate the CE-BW for all $\lambda_p$ and $W_{wg}$ when the $\Lambda$ is quasi phase-matched for degenerate DFG. We calculate the CE-BW of the "middle" or "distant" peaks if the respective peak's -3dB intensity level does not cross the other peak. We label the data points that are plotted in Fig.~\ref{Fig3}c with red markers. In the "middle peak" ("distant peak"), CE-BW can be tuned from $\sim$200~nm~($\sim$40~nm) to $\sim$600~nm~($\sim$220~nm) by precisely selecting $W_{wg}$ as shown in Fig.~\ref{Fig3}d~(Fig.~\ref{Fig3}e). CE-BW of both "middle peak" and "distant peak" increases with increasing $W_{wg}$ for all $\lambda_p$. We additionally calculate CE-BW of the "joint peak" when the CE-BW cannot be individually determined for neither "middle peak" nor "distant peak". We observe the broadest bandwidth right after the peaks merger, which can exceed 780~nm.  Overall, within this entire design space, it is possible to have a tunable CE-BW covering the telecom bands from $\sim$200~nm to $\sim$780~nm near the degenerate DFG regime by controlling $W_{wg}$ .

To summarize, in this section, we have presented a detailed analysis of optimizing $\Lambda$, $\lambda_{p}$ and the waveguide cross-section to either achieve highest CE-BW or to precisely tune CE-BW. We obtain maximum CE-BW of $\sim$300~nm in the cladded configuration and $\sim$780~nm in the uncladded configuration. To our knowledge, this analysis provides one of the widest bandwidths foreseen for any TFLN waveguide relying on DFG \cite{Wei2024EfficientWaveguides, Javid2021UltrabroadbandChip, Xue2021UltrabrightChip, Fang2024EfficientNanowaveguides, Sua2018Ultra-widebandWaveguide, Ledezma2022IntenseWaveguides, Mishra2022Ultra-broadbandNiobate}.

\section{Optical Characterization of Fabricated Waveguides}
\label{sec3:meas}

We experimentally verify our analysis on one of the favorable broadband cladded cross-sections with $W_{wg}$~=~1200~nm~(Fig.~\ref{Fig1}a). Several waveguides with these nominal dimensions and 4.8~mm PPLN length were fabricated at CSEM in a die size of 5x5 mm$^{2}$ on different wafers. Fabrication of the waveguide and electrode follows standard process available in the TFLN PIC foundry \cite{Sattari2024StandardizedPlatformb}. Here the electrodes are deposited after waveguide patterning and located in the standard metal layers. They have square tips with 50\% duty cycle. For poling, we apply a series of 0.4 ms pulses from these electrodes. To take into account possible wafer to wafer fluctuations, we patterned electrodes with different $\Lambda$~(namely 4100~nm, 4200~nm and 4300~nm). 

% Description of Setups:
The setup used for optical characterization is shown in Fig.~\ref{Fig4}a. The chip is placed on a temperature controlled stage that is constructed from a PID controller and a Peltier element. Different tunable telecom lasers~(Yenista Tunics T100) and an NIR laser~(Newfocus TLB-6700) are used to sweep $\lambda_s$ and $\lambda_p$, respectively. For $\lambda_p$ sweeps, we aim to observe both the generated idler peak and its sidelobes for a selected $\lambda_s$. In this case, we amplify a fixed telecom signal with an EDFA and reduce the noise floor by placing a band-pass filter (BPF) after the amplifier. We control the polarization of both inputs and combine them after isolators with a wavelength-division multiplexer~(WDM). We use lensed fibers (cone angle~=~90$^{\circ}$ $\pm$ 5$^{\circ}$) for coupling in and out from the chip. We record the outcoupled pump, signal and idler spectra in an optical spectrum analyser (OSA Yokogawa AQ6370). For $\lambda_s$ sweeps at a fixed $\lambda_p$, we optimize the setup for sweeping a wide wavelength range from 1250~nm to 1660~nm while minimizing losses. Consequently, we remove the EDFA, band-pass filter, isolator and NIR polarization controller from the signal path. For outcoupling the broadband light from the chip, we use an objective~(NA~=~0.35) followed by a reflective collimator which delivers light to the  OSA~(Yokogawa AQ6375). 

Prior to DFG characterization, we also measure the SH-response in the same setup by coupling only the telecom laser into the chip from a lensed fiber. We outcouple the light with the objective and separate telecom-input and generated-SH with a dichroic mirror~(DMSP). With this technique, we simultaneously detect telecom-input and generated SH from individual powermeters (PM).

\subsection{Effective Poling Period on Fabricated Waveguide}
\label{sec3.1:shg&dfg-pump}

\begin{figure}  [!htb]
\centering\includegraphics[width=13cm]{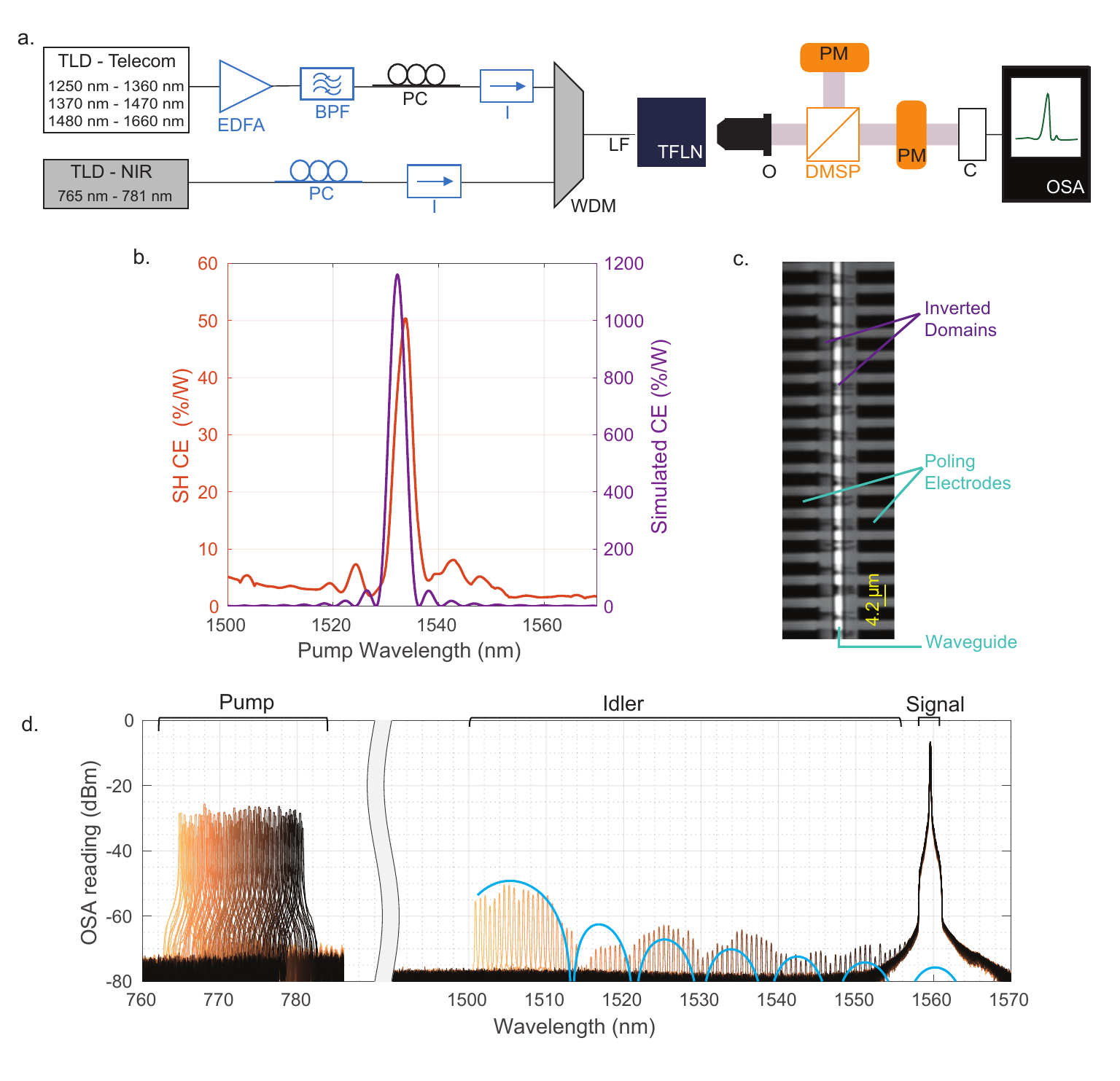}
\caption{a) Optical characterization setup. Light blue components are only used in DFG pump sweep. Orange components are only used in SHG characterization. TLD:~tunable laser diode, PC:~polarization controller, I:~isolator, EDFA:~erbium-doped
fiber amplifier, BPF:~band-pass filter, WDM:~wavelength division multiplexer, LF:~lensed
fiber, O:~objective, DMSP:~short-pass dichroic mirror, PM:~powermeter, C:~collimator, OSA:~optical spectrum analyser. b)~Measured SH CE for WG-1 is shown in orange axes, simulated SHG CE for poling period $\Lambda$~=~4241~nm is shown in purple axes. c) TPM image of WG-1 after poling. d) OSA traces showing the swept pump~(from 765~nm to 781~nm), signal~($\sim$1560~nm) and generated idler in WG-1. Simulated DFG CE~(blue line) is overlaid on the measurements in arbitrary units.}
\label{Fig4}
\end{figure}

We perform SH measurements and identify a waveguide~(WG-1) with the SH-peak at $\sim$1533~nm corresponding to $\Lambda$~=~$\sim$4241~nm for the simulated design cross-section. In Fig.~\ref{Fig4}b, we plot the measured SH CE~(calculated as $P_\textnormal{SH}/(P_\textnormal{Pump})^2$) for WG-1 and overlay with the simulations for $\Lambda$~=~4241~nm as it differs from the actual poling period because of fabrication tolerances in the waveguide dimensions\cite{Koyaz2024DesignWaveguides} (discussed in detail in Sec.~\ref{sec4:conc}). We refer to the $\Lambda$~=~4241~nm as the effective poling period of WG-1 as it facilitates the measured response at the targeted design.

We report an acceptable agreement in SH-peak position between simulations~($\sim$1532.2~nm) and measurements~($\sim$1533.8~nm). Yet the measured peak intensity is noticeably lower, which we mainly associate with the imperfections on the poling quality. The quality of periodic poling is inspected using two-photon microscopy~(TPM). In the TPM image~(Fig.~\ref{Fig4}c), the inverted domains appear darker compared to the unpoled regions and thus indicate underpoling, which can result in a reduced CE. We also recognize the potential role of height fluctuations along the waveguide in the reduced CE~\cite{Chen2023AdaptedWaveguides, Zhao2023UnveilingDoublers} (further discussed in Sec.~\ref{sec3.2:broad-dfg}\&\ref{sec4:conc}).

After SHG measurements, we characterize DFG in a pump sweep for $\lambda_s$=1560~nm and $\lambda_p$ ranging from 765~nm to 781~nm. OSA traces are shown in Fig.~\ref{Fig4}d. We obtain a decent agreement between simulations~(blue line) and measurements especially on the main peak. We associate the discrepancies on the sidelobes with the imperfections on poling quality and the inhomogeneities in waveguide dimensions. Additionally, we observe prominent wavelength dependent oscillations on the pump and the idler~(the origin of these oscillations will be discussed in Sec.~\ref{sec4:conc}). Still, with these two sets of independent measurements~(SH characterization and DFG pump sweep), we confirm $\Lambda$~=~$\sim$4241~nm as the effective poling period of WG-1.

\subsection{Broadband DFG and Further Tuning Mechanisms}
\label{sec3.2:broad-dfg}

\begin{figure}   [!htb]
\centering\includegraphics[width=13cm]{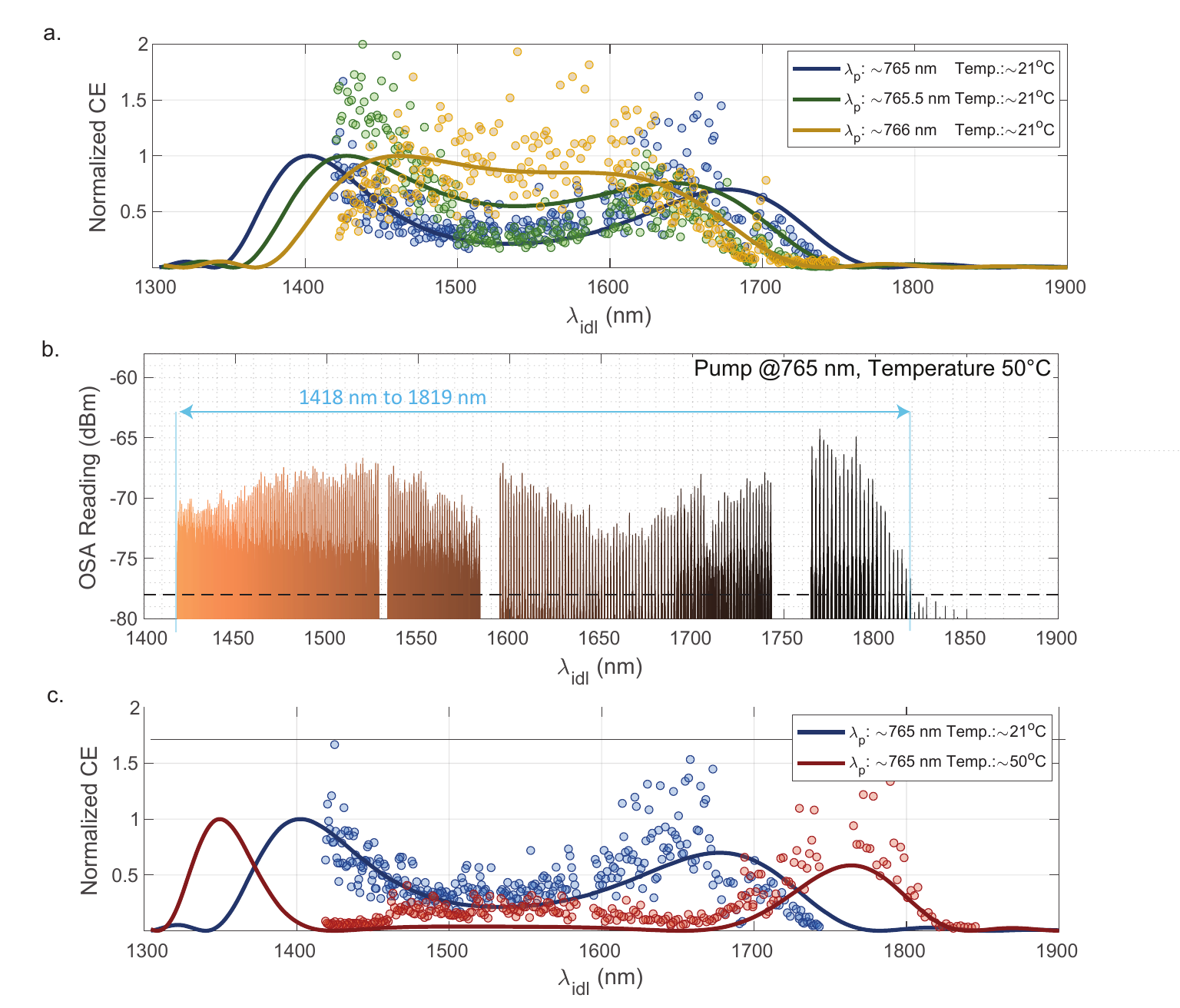}
\caption{a) Normalized DFG CE of $\lambda_{s}$ sweep for WG-1 at different $\lambda_{p}$ namely 765~nm, 765.5~nm and 766~nm, b) Superimposed OSA traces showing the generated idler at stage temperature 50°C, c) Normalized DFG CE of $\lambda_{s}$ sweep for the same waveguide at different temperatures~(21°C and 50°C) for $\lambda_{p}$~=~765~nm. Measurements in data points, simulated CE in solid lines in both (a) and (c). Simulation parameter $\Lambda$ is 4241~nm at stage temperature 21°C and 4244~nm at stage temperature 50°C. }
\label{Fig5}
\end{figure}

\begin{figure}    [!htb]
\centering\includegraphics[width=13cm] {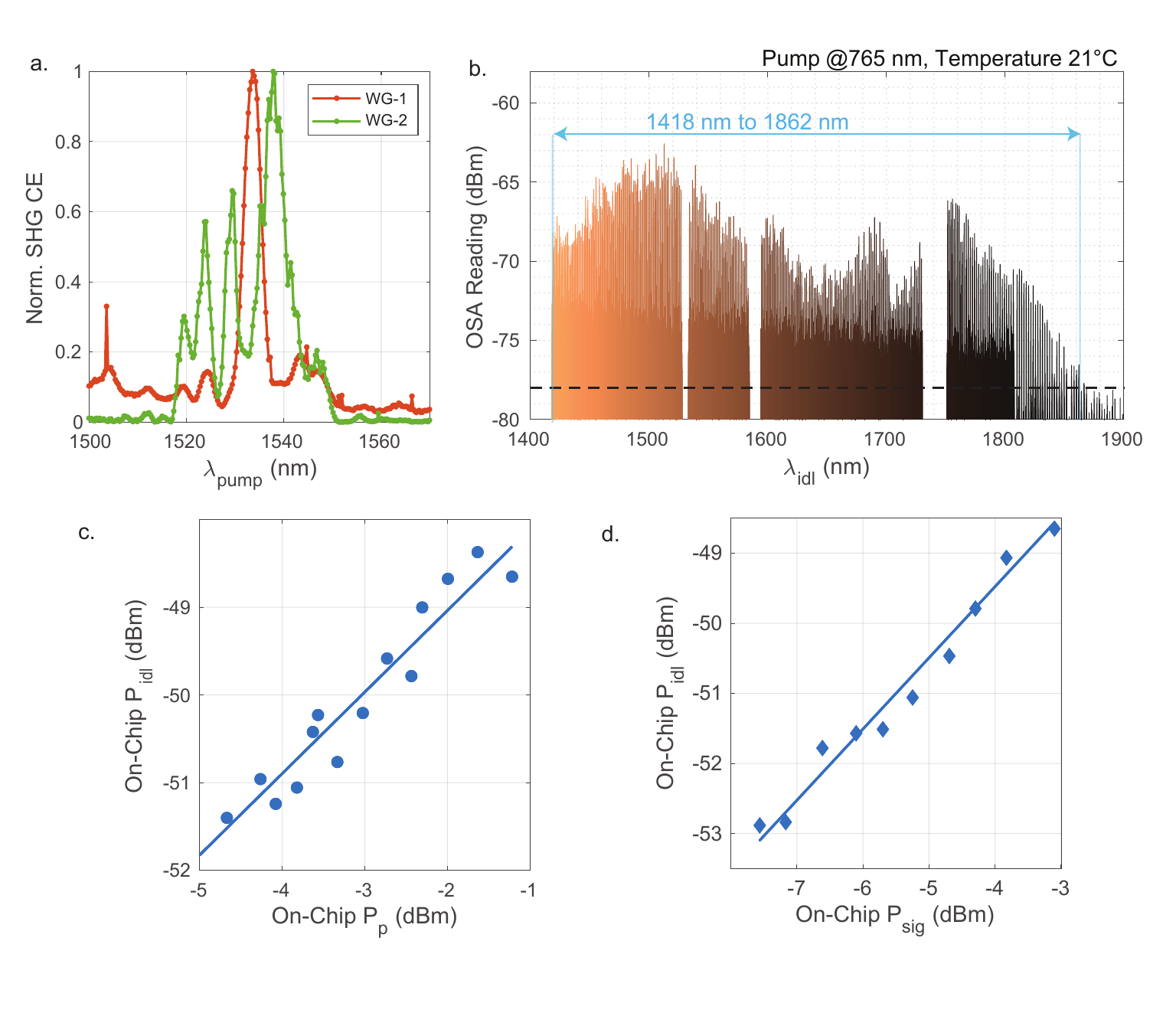}
\caption{a) Normalized SHG CE of WG-1 and WG-2. b) OSA traces showing the generated idler on WG-2 while sweeping the signal for a pump at 765~nm. c,~d) Generated on-chip idler power while sweeping c.~pump and d.~signal powers, for 765~nm pump and 1550~nm signal.  }
\label{Fig6}
\end{figure}

To verify the simulated broadband response of WG-1, we sweep $\lambda_{s}$ at three $\lambda_{p}$~(namely 765~nm, 765.5~nm and 766~nm) and record the DFG response on the OSA using the setup illustrated in Fig.~\ref{Fig5}a. We obtain the total signal and idler powers by integrating the spectral components throughout the measured spectra and calculate the on-chip power by taking wavelength dependent coupling losses into account. 

To extract coupling losses, we perform additional transmission measurements for input laser wavelengths~($\lambda_{las}$) ranging from 1260 nm to 1660 nm
, for $\lambda_{las}$$\sim$2~$\mu$m and for $\lambda_{las}$ ranging from 765 nm to 781 nm when the chip is placed on~(and removed from) the stage.  We post-process these transmission measurements by using a moving average filter to smoothen the data from the wavelength-dependent oscillations (further elaborated in Sec.~\ref{sec4:conc}).

Using this technique, we obtain measured-CE~(calculated as $P_\textnormal{i} / (P_\textnormal{s} P_\textnormal{p})$) as a function of $\lambda_{i}$ (analogous to $\lambda_{s}$) for different pump wavelengths ($\lambda_{p}$) namely 765 nm, 765.5 nm, and 766 nm. To compare the observed idler generation wavelength range with simulations, we fit the simulated CE to the measured data using Eq.~\ref{eq-ce}. In this fitting process, the amplitude serves as the fitting parameter, while $\lambda_{p}$ and the effective poling period (determined in Sec.~\ref{sec3.1:shg&dfg-pump}) are simulation parameters. In Fig.~\ref{Fig5}a, the measured CE (data points) and the fitted CE (solid lines) are shown after normalization to the maximum value of the fit for each corresponding $\lambda_{p}$. The measurements are in accordance with the simulations, showing idler generation from 1418~nm to 1740~nm, limited by our available sources. We also observe the separation of the phased-matched regions as the pump is tuned from 766~nm to 765~nm. 

Furthermore, we perform another measurement at $\sim$50°C whose superimposed OSA spectra is shown in Fig.~\ref{Fig5}b. In this case, idler generation up to $\lambda_{i}$~=~1819~nm is measured.  We also plot the normalized CE as a function of $\lambda_{i}$ at different temperatures~(namely 21°C and 50°C) as shown in Fig.~\ref{Fig5}c. In our analysis, this temperature range is equivalent to a 3~nm period shift (which lies within the tuning range that can be facilitated by the thermo-optic effect on LN ~\cite{Fejer1992Quasi-phase-matchedTolerances, Ghosh1994Thermo-opticCrystals, Luo2018HighlyWaveguide}) and corresponds to effective poling period of 4244~nm. We demonstrate that idler generation shifts towards mid-IR with increasing temperature. 

We then investigate the impact of longitudinal waveguide fluctuations on idler generation. For this purpose, we characterize another waveguide~(WG-2) with the same nominal design cross section as WG-1 but with a different SH response as shown in Fig.~\ref{Fig6}a. The SH spectrum of WG-2 has three distinct peaks indicating the presence of fluctuations in the waveguide cross section along the propagation direction due to TFLN thickness variations on the wafer and etching process uncertainties. We estimate the fluctuation in the total TFLN height to be $>$2~nm~(approximately twice that of WG-1) and the fluctuation of the slab thickness to be around $\sim$10\%~($\sim$35\% larger than WG1), based on TFLN substrate's datasheet and ellipsometer measurements performed by CSEM. We calculate the poling period within the presence of mentioned fluctuations along the waveguide. These calculations indicate a shift of tens of nanometres in $\Lambda$ between the two ends of the waveguide. Wu et al.~(2022) demonstrate that a small perturbation like chirping~(with a range at around $\sim$50~nm) can significantly increase the SHG bandwidth~\cite{Wu2022BroadbandWaveguides}. Here, we experimentally show the broadened DFG response due to longitudinal waveguide fluctuations for a comparable perturbation on $\Lambda$. In WG-2, we observe the DFG idler from 1418~nm to 1862~nm, when injecting $\lambda_{s}$ from 1660~nm to 1294~nm at a fixed temperature~($\sim$21°C) and $\lambda_p$~(765~nm) (Fig. \ref{Fig6}b). This indicates that idler generation is possible from 1294~nm to 1862~nm.

In addition to our main focus, i.e. broadband process,  we measured the CE and its dependence on input power in WG-2. This value could be improved in future work with the optimization of the poling quality, as previously stated. To accurately estimate the on-chip power from OSA traces, we scale them with powermeter readings~(from Thorlabs S120C and S148C) to account for any calibration loss between the collimator and the OSA input. With this technique, we obtain CE between 4-7~\%/W at $\lambda_{s}$~=~1550~nm and $\lambda_{p}$~=~765~nm. Finally, we sweep the on-chip pump~(signal) power and record on-chip idler~(idler) power as shown in Fig.~\ref{Fig6}c(d) while keeping $\lambda_{s}$~=~1550~nm and $\lambda_{p}$~=~765~nm constant. A first-degree polynomial fit is then applied to the data, yielding slopes of 0.9 (for the pump) and 1.0 (for the signal) for the idler power as a function of the pump (signal) inputs. The slight deviation from the ideal slope for the pump can be attributed to wavelength-dependent oscillations (further discussed in Sec. 4.)   All in all,  this analysis confirms the expected linear dependence (i.e., a slope of 1) of signal and pump inputs.

\section{Discussion and Conclusion}
\label{sec4:conc}

\begin{table}[h!]
\centering 
\scriptsize 
\begin{tabular}{|>{\centering\arraybackslash}p{1.5cm}>{\centering\arraybackslash}p{1.5cm}>{\centering\arraybackslash}p{1.5cm}>{\centering\arraybackslash}p{2.7cm}>{\centering\arraybackslash}p{2.3cm}>{\centering\arraybackslash}p{1.5cm}|} 
\hline 
\rowcolor{gray!50}
  & Focus & $\lambda_{p}$ (nm) & Meas. $\lambda_{idl}$ (nm) & Sim. $\lambda_{idl}$ (nm) & Length (mm)\\  
\hline
\rule{0pt}{12pt}
this work & DFG & $\sim$765 & \parbox{2cm}{\centering WG1: 1418-1819$^{(2)}$ \\ WG2: 1418-1862$^{(2)}$ } & 1160-2250 & 4.8-5 \\ 
\rule{0pt}{0pt}
 &  &  &  &  & \\
\rule{0pt}{12pt}
\cite{Wei2024EfficientWaveguides} & C-O WC & $\sim$1550 & \parbox{2cm}{\centering 1290-1360 and \\ 1500-1600$^{(2)}$} & 1250-1650 & 5 \\ 
\rule{0pt}{12pt}
\cite{Javid2021UltrabroadbandChip} & SPDC & $\sim$770 & 1200-1600$^{(1)}$ & 1200-2000 & 5 \\
\rule{0pt}{12pt}
\cite{Xue2021UltrabrightChip} & SPDC & $\sim$735 & 1350-1600 & 1350-1600 & 6  \\
\rule{0pt}{12pt}
\cite{Fang2024EfficientNanowaveguides} & SPDC & $\sim$810 & 1500-1620$^{(1)}$ & 1400-1800 & 5.7  \\
\rule{0pt}{12pt}
\cite{Niu2020OptimizingDomains} & SHG/DFG & $\sim$735 & 1524-1581 & n.a. & 6   \\
\rule{0pt}{12pt}
\cite{Sua2018Ultra-widebandWaveguide} & OPA & $\sim$1550 & 1495-1609 & 1475-1620 & 20 \\ 
\rule{0pt}{12pt}
\cite{Ledezma2022IntenseWaveguides} & OPA & $\sim$1045 & 1600-2400 $^{(3)}$ & 1600-2800 & 6  \\
\rule{0pt}{12pt}
\cite{Mishra2022Ultra-broadbandNiobate} & DFG & $\sim$1064 & 2800–3800 & 2800–3800 & 5  \\ 
\rule{0pt}{2pt}
&  &  &  &  & \\
\hline 

\multicolumn{6}{|c|}{\parbox{13cm}{
\vspace{6pt}  \scriptsize $\lambda_{p}$: Pump Wavelength, Meas. $\lambda_{idl}$: Wavelength range of experimentally generated idler, Sim. $\lambda_{idl}$: Wavelength range of generated idler in simulations, Length: Length of periodically poled waveguide (z)\\
\scriptsize \\
\scriptsize  C-O WC: C-Band to O-Band Wavelength Conversion, SPDC : Spontaneous Parametric Down Conversion, 
\scriptsize  OPA: Optical Parametric Amplification, 
$^{(1)}$ limited with photodetection range, $^{(2)}$ limited by available laser sources, $^{(3)}$ limited by OSA detection range,
}} \\
\hline
\end{tabular}
\caption{Comparison of DFG in periodically-poled TFLN}
\label{tab:comparison} 
\end{table}

In this work, we focus on realization of broadband DFG frequency converters based on  CSEM's TFLN foundry stacks. We deem this as an essential step in view of implementing nonlinear PDK building blocks for large scale TFLN PICs.

We successively optimize $\Lambda$, $\lambda_{p}$ and $W_{wg}$ in different dispersion regimes, namely normal, near-zero and anomalous. For broadband operation, we show that the broadest bandwidth occurs on near-zero dispersion, and we achieve a 3-dB DFG CE bandwidth of $\sim$300~nm in cladded configuration and $\sim$780~nm in the uncladded configuration through dispersion engineering.

We overlay the results of simulations with measurements by extracting the effective poling period from the SH-peak and obtain a decent agreement with measurements. As the electrode periodicity is defined lithographically and has negligible fabrication error, we associate the discrepancy between this periodicity and the effective poling period with the fabrication deviations from the designed waveguide cross-section~\cite{Koyaz2024DesignWaveguides, Chen2023AdaptedWaveguides, Zhao2023UnveilingDoublers} and the use of material data from literature~ \cite{Zelmon1997InfraredNiobate}. These deviations could modify the phase-matched poling period and consequently, shift the optical response. In our methodology, these are considered as perturbations on the design and accounted for in DFG measurements by using the effective poling period. We emphasize on the consistency of the effective poling period~($\Lambda$~=~4241~nm on WG1) in all SHG and DFG measurements with pump and signal sweeps. With this approach, we provide a method to bridge the gap between the targeted design and the characterized waveguide cross-section.

In the measurements, we observe strong wavelength-dependent oscillations possibly due to Fabry-Perot resonances or due to the presence of higher-order modes. Firstly, to eliminate Fabry-Perot resonances caused by facet reflectivity, we tested different coupling configurations. However, none of these configurations fully suppressed the oscillations, reducing the likelihood of Fabry-Perot as the main origin of the observed oscillations. Alternatively, the higher-order modes at the pump and signal wavelengths originate from the input coupling scheme, as the fiberized WDM and lensed fiber are multi-mode for the pump. These could also couple into the chip since the investigated waveguides do not have any dedicated tapers. For these waveguides, we numerically calculate a total of 4 optical modes at 1550 nm and 11 optical modes at 765 nm in simulations. As a result, it is probable that a portion of light will couple in a different mode from the targeted fundamental TE mode. Such fluctuations in CE have been studied in literature for different platforms, like SiN, and other photon mixing processes like FWM \cite{Ayan2023TowardsWaveguides}. As a result, we associate the presence of higher-order modes at the pump and signal wavelengths as the main cause of these oscillations. They can be minimized by implementing dedicated taper designs. Despite these oscillations, there is a clear agreement between our measurements and simulations.

We also recognize that our measured CE is noticeably lower compared to simulated values. We primarily attribute this with the poling quality. TPM images of the investigated waveguides suggest underpoling (as discussed in Sec.~\ref{sec3.1:shg&dfg-pump}). Since we perform poling after waveguide patterning, the poled domains start to form on the slab and primarily grow laterally along the slab\cite{Zhao2019OpticalWaveguides}. As a result, part of the waveguide ridge could remain underpoled even if there is complete inversion on the slab. According to our simulations, when 50\% of the total waveguide height is inverted, the peak conversion efficiency is reduced by a factor of $\sim$3.8. In our case, the slab is thinner, potentially leading to an even more significant reduction in peak CE. Additionally, the presence of longitudinal variations on waveguide cross-section can limit peak CE, especially on WG-2. In the case of broadband designs, we show that the longitudinal waveguide fluctuations are equivalent to chirping of the poling period. Regardless, we perform the broadband measurements without requiring to any additional optical amplifier in the beam path.

We experimentally obtain DFG idler from 1418~nm to 1740~nm and further broaden the range of idler generation via temperature tuning~(up to $\lambda_{i}$~=~1819~nm) and by leveraging the longitudinal waveguide fluctuations~(up to $\lambda_{i}$~=~1862~nm), thereby extending the operation towards IR. We focus our study on both cladded and uncladded waveguides, with the former being experimentally verified. While we show that uncladded waveguides can offer additional benefit in view of flexible and large bandwidth designs, such approach could require additional optimization in terms of fabrication and poling after waveguide patterning due to increased surface currents affecting the poled domain formation~\cite{Mizuuchi2004Electric-fieldLiNbO3, Nagy2019ReducingFilms}.

Compared to prior work on periodically-poled TFLN~(Table.~1), our study demonstrates DFG bandwidth improvement both in terms of designs and experiments. We present a core building block of nonlinear optics capable of generating signal and idler wavelengths across the entire telecommunication bands. We acknowledge its potential functionality in a multitude of applications including multichannel wavelength-division multiplexing, broadband comb generation and cascaded supercontinuum generation, all while leveraging its availability in a wafer-scale industrial platform.

\begin{backmatter}
\bmsection{Funding}
 Schweizerischer Nationalfonds zur Förderung der Wissenschaftlichen Forschung (203480).

\bmsection{Acknowledgments}
 The authors acknowledge CSEM cleanroom staff for manufacturing the TFLN chips. We also thank F. Ayhan for valuable help on TFLN poling and N. Balakleyskiy for fruitful discussions on design.
 % This work is supported by the Swiss National Science Foundation (SNSF) Bridge Discovery project (40B2-0) 203480, ENABLE.

\bmsection{Disclosures}
The authors declare that there are no conflicts of interest related to this article.

\bmsection{Data Availability}
Data underlying the results presented in this paper are not publicly available at this time but may be obtained from the authors upon reasonable request.

\end{backmatter}

\bibliography{dfg_references}

\begin{thebibliography}{10}
\newcommand{\enquote}[1]{``#1''}

\bibitem{Liu2005NonlinearDevices}
J.-M. Liu, \enquote{{Nonlinear Optical Devices},} in \emph{Photonic Devices,}  (Cambridge University, 2005), pp. 441--610.

\bibitem{Nassau1966FerroelectricCrystals}
K.~Nassau, H.~J. Levinstein, and G.~M. Loiacono, \enquote{{Ferroelectric lithium niobate. 2. Preparation of single domain crystals},} {\protect\JournalTitle{J. Phys. Chem. Solids}} \textbf{27}, 989--996 (1966).

\bibitem{Nassau1966FerroelectricEtching}
K.~Nassau, H.~J. Levinstein, and G.~M. Loiacono, \enquote{{Ferroelectric lithium niobate. 1. Growth, domain structure, dislocations and etching},} {\protect\JournalTitle{J. Phys. Chem. Solids}} \textbf{27}, 983--988 (1966).

\bibitem{Wang2017SecondWaveguides}
C.~Wang, X.~Xiong, N.~Andrade, \emph{et~al.}, \enquote{{Second harmonic generation in nano-structured thin-film lithium niobate waveguides},} {\protect\JournalTitle{Opt. Express}} \textbf{25}, 6963--6973 (2017).

\bibitem{Wu2018LongRoughness}
R.~Wu, M.~Wang, J.~Xu, \emph{et~al.}, \enquote{{Long Low-Loss-Litium Niobate on Insulator Waveguides with Sub-Nanometer Surface Roughness},} {\protect\JournalTitle{Nanomaterials}} \textbf{8}, 910 (2018).

\bibitem{Zhu2021IntegratedNiobate}
D.~Zhu, L.~Shao, M.~Yu, \emph{et~al.}, \enquote{{Integrated photonics on thin-film lithium niobate},} {\protect\JournalTitle{Adv. Opt. Photon.}} \textbf{13}, 242--352 (2021).

\bibitem{Cerullo2003UltrafastAmplifiers}
G.~Cerullo and S.~De~Silvestri, \enquote{{Ultrafast optical parametric amplifiers},} {\protect\JournalTitle{Review of Scientific Instruments}} \textbf{74}, 1--18 (2003).

\bibitem{Zhang2019OpticalAmplification}
J.-Y. Zhang, \enquote{{Optical Parametric Generation and Amplification},} in \emph{Optical Parametric Generation and Amplification,}  (Routledge, 2019), chap.~2, 1st ed.

\bibitem{Mishra2022Ultra-broadbandNiobate}
J.~Mishra, M.~Jankowski, A.~Hwang, \emph{et~al.}, \enquote{{Ultra-broadband mid-infrared generation in dispersion-engineered thin-film lithium niobate},} {\protect\JournalTitle{Opt. Express}} \textbf{30}, 32752--32760 (2022).

\bibitem{Ledezma2022IntenseWaveguides}
L.~Ledezma, R.~Sekine, Q.~Guo, \emph{et~al.}, \enquote{{Intense optical parametric amplification in dispersion-engineered nanophotonic lithium niobate waveguides},} {\protect\JournalTitle{Optica}} \textbf{9}, 303--308 (2022).

\bibitem{Niu2020OptimizingDomains}
Y.~Niu, C.~Lin, X.~Liu, \emph{et~al.}, \enquote{{Optimizing the efficiency of a periodically poled LNOI waveguide using in situ monitoring of the ferroelectric domains},} {\protect\JournalTitle{Appl. Phys. Lett.}} \textbf{116}, 101104 (2020).

\bibitem{Sua2018Ultra-widebandWaveguide}
Y.~M. Sua, J.-Y. Chen, and Y.-P. Huang, \enquote{{Ultra-wideband and high-gain parametric amplification in telecom wavelengths with an optimally mode-matched PPLN waveguide},} {\protect\JournalTitle{Opt. Lett.}} \textbf{43}, 2965--2968 (2018).

\bibitem{Kishimoto2016HighlyWaveguide}
T.~Kishimoto, K.~Inafune, Y.~Ogawa, \emph{et~al.}, \enquote{{Highly efficient phase-sensitive parametric gain in periodically poled LiNbO3 ridge waveguide},} {\protect\JournalTitle{Opt. Lett.}} \textbf{41}, 1905--1908 (2016).

\bibitem{Wei2024EfficientWaveguides}
J.~Wei, C.~Cheng, Y.~Wu, \emph{et~al.}, \enquote{{Efficient and Broadband All-Optical Wavelength Conversion between C and O Bands in Nanophotonic Lithium Niobate Waveguides},} {\protect\JournalTitle{Journal of Lightwave Technology}} \textbf{42}, 5966--5973 (2024).

\bibitem{Fang2024EfficientNanowaveguides}
X.-X. Fang, L.~Wang, and H.~Lu, \enquote{{Efficient generation of broadband photon pairs in shallow-etched lithium niobate nanowaveguides},} {\protect\JournalTitle{Opt. Express}} \textbf{32}, 22945--22954 (2024).

\bibitem{Javid2021UltrabroadbandChip}
U.~A. Javid, J.~Ling, J.~Staffa, \emph{et~al.}, \enquote{{Ultrabroadband Entangled Photons on a Nanophotonic Chip},} {\protect\JournalTitle{Phys. Rev. Lett.}} \textbf{127}, 183601 (2021).

\bibitem{Xue2021UltrabrightChip}
G.-T. Xue, Y.-F. Niu, X.~Liu, \emph{et~al.}, \enquote{{Ultrabright Multiplexed Energy-Time-Entangled Photon Generation from Lithium Niobate on Insulator Chip},} {\protect\JournalTitle{Phys. Rev. Applied}} \textbf{15}, 064059 (2021).

\bibitem{Suhara2003WaveguideDevices}
T.~Suhara and M.~Fujimura, \enquote{{Waveguide Nonlinear-Optic Devices},}  (Springer Berlin, 2003), pp. 193--270, 1st ed.

\bibitem{Sutherland2003HandbookOptics}
R.~L. Sutherland, \enquote{{Handbook of Nonlinear Optics},}  (CRC Press, 2003), pp. 33--85, 2nd ed.

\bibitem{Sattari2024StandardizedPlatformb}
H.~Sattari, I.~Prieto, H.~Zarebidaki, \emph{et~al.}, \enquote{{Standardized TFLN Photonic Integrated Circuits Platform},} in \emph{The 25th European Conference on Integrated Optics.},  J.~Witzens, J.~Poon, L.~Zimmermann, and W.~Freude, eds. (Springer Nature Switzerland, 2024), pp. 85--89.

\bibitem{Pu2018UltraEfficientProcessing}
M.~Pu, H.~Hu, L.~Ottaviano, \emph{et~al.}, \enquote{{Ultra‐Efficient and Broadband Nonlinear AlGaAs‐on‐Insulator Chip for Low‐Power Optical Signal Processing},} {\protect\JournalTitle{Laser {\&} Photonics Reviews}} \textbf{12}, 1800111 (2018).

\bibitem{Koyaz2024DesignWaveguides}
Y.~Koyaz, C.~Lafforgue, N.~Balakleyskiy, \emph{et~al.}, \enquote{{Design rules for frequency conversion in periodically poled thin film lithium niobate waveguides},} in \emph{Integrated Optics: Devices, Materials, and Technologies XXVIII,}  vol. 12889 (SPIE, 2024), p. 128890O.

\bibitem{Chen2023AdaptedWaveguides}
P.~K. Chen, I.~Briggs, C.~Cui, \emph{et~al.}, \enquote{{Adapted poling to break the nonlinear efficiency limit in nanophotonic lithium niobate waveguides},} {\protect\JournalTitle{Nat. Nanotechnol.}} \textbf{19}, 44--50 (2023).

\bibitem{Zhao2023UnveilingDoublers}
J.~Zhao, X.~Li, T.~C. Hu, \emph{et~al.}, \enquote{{Unveiling the origins of quasi-phase matching spectral imperfections in thin-film lithium niobate frequency doublers},} {\protect\JournalTitle{APL Photonics}} \textbf{8}, 126106 (2023).

\bibitem{Fejer1992Quasi-phase-matchedTolerances}
M.~Fejer, G.~Magel, D.~Jundt, and R.~Byer, \enquote{{Quasi-phase-matched second harmonic generation: tuning and tolerances},} {\protect\JournalTitle{IEEE Journal of Quantum Electronics}} \textbf{28}, 2631--2654 (1992).

\bibitem{Ghosh1994Thermo-opticCrystals}
G.~Ghosh, \enquote{{Thermo-optic coefficients of LiNbO3, LiIO3, and LiTaO3 nonlinear crystals},} {\protect\JournalTitle{Opt. Lett.}} \textbf{19}, 1391--1393 (1994).

\bibitem{Luo2018HighlyWaveguide}
R.~Luo, Y.~He, H.~Liang, \emph{et~al.}, \enquote{{Highly tunable efficient second-harmonic generation in a lithium niobate nanophotonic waveguide},} {\protect\JournalTitle{Optica}} \textbf{5}, 1006--1011 (2018).

\bibitem{Wu2022BroadbandWaveguides}
X.~Wu, L.~Zhang, Z.~Hao, \emph{et~al.}, \enquote{{Broadband second-harmonic generation in step-chirped periodically poled lithium niobate waveguides},} {\protect\JournalTitle{Opt. Lett.}} \textbf{47}, 1574--1577 (2022).

\bibitem{Zelmon1997InfraredNiobate}
D.~E. Zelmon, D.~L. Small, and D.~Jundt, \enquote{{Infrared corrected Sellmeier coefficients for congruently grown lithium niobate and 5 mol. {\%} magnesium oxide–doped lithium niobate},} {\protect\JournalTitle{J. Opt. Soc. Am. B}} \textbf{14}, 3319--3322 (1997).

\bibitem{Ayan2023TowardsWaveguides}
A.~Ayan, J.~Liu, T.~J. Kippenberg, and C.-S. Br{\`{e}}s, \enquote{{Towards efficient broadband parametric conversion in ultra-long Si3N4 waveguides},} {\protect\JournalTitle{Opt. Express}} \textbf{31}, 40916--40927 (2023).

\bibitem{Zhao2019OpticalWaveguides}
J.~Zhao, M.~R{\"{u}}sing, and S.~Mookherjea, \enquote{{Optical diagnostic methods for monitoring the poling of thin-film lithium niobate waveguides},} {\protect\JournalTitle{Opt. Express}} \textbf{27}, 12025--12038 (2019).

\bibitem{Mizuuchi2004Electric-fieldLiNbO3}
K.~Mizuuchi, A.~Morikawa, T.~Sugita, and K.~Yamamoto, \enquote{{Electric-field poling in Mg-doped LiNbO3},} {\protect\JournalTitle{J. Appl. Phys.}} \textbf{96}, 6585--6590 (2004).

\bibitem{Nagy2019ReducingFilms}
J.~T. Nagy and R.~M. Reano, \enquote{{Reducing leakage current during periodic poling of ion-sliced x-cut MgO doped lithium niobate thin films},} {\protect\JournalTitle{Opt. Mater. Express}} \textbf{9}, 3146--3155 (2019).

\end{thebibliography}

\end{document}